\documentclass[fleqn,twoside]{article}
\usepackage{espcrc2,natbib}

\usepackage{graphicx}

\def\myputfigure#1#2#3#4#5%
{\vskip#5pt\makebox[0pt]{\hskip#2in
\includegraphics[width=#3\textwidth]{#1}}\vskip#4pt\hfill}

\title{Studying Dark Energy with Galaxy Cluster Surveys}

\author{Joseph J. Mohr\address[Illinois]{University of Illinois Departments of Astronomy and of Physics,
        Urbana, IL USA}\thanks{A special thanks to survey science collaborators Gil Holder 
	and Subha Majumdar for stimulating conversations.  This work was supported in part by NASA Long
	Term Space Astrophysics award NAG5-11415 and {\it Chandra} X--ray Observatory archival grant
	AR1-2002X.},
	Brian O'Shea\addressmark[Illinois],
	August E. Evrard\address[Michigan]{University of Michigan Departments of Physics and Astronomy,
	Ann Arbor, MI  USA},
	John Bialek\addressmark[Michigan], 
	Zoltan Haiman\address{Princeton University Observatory, Princeton, NJ USA}
	}
	
\begin{document}

\begin{abstract}
Galaxy cluster surveys provide a powerful means of studying the density and nature of the dark energy.  The redshift distribution of detected clusters in a deep, large solid angle SZE or X--ray survey is highly sensitive to the dark energy equation of state.  Accurate constraints at the 5\% level on the dark energy equation of state require that systematic biases in the mass estimators must be controlled at  better than the $\sim$10\% level.    Observed regularity in the cluster population and the availability of multiple, independent mass estimators suggests these precise measurements are possible.  Using hydrodynamical simulations that include preheating, we show that the level of preheating required to explain local galaxy cluster structure has a dramatic effect on X--ray cluster surveys, but only a mild effect on SZE surveys.  This suggests that SZE surveys may be optimal for cosmology while X--ray surveys are well suited for studies of the thermal history of the intracluster medium.
\vspace{1pc}
\end{abstract}

\maketitle

\section{Cluster Surveys and Cosmology}
Galaxy clusters have long been used to study dark matter and cosmology in general.  Cluster surveys in the local universe are particularly useful for constraining a combination of the matter density parameter $\Omega_M$ and the normalization of the power spectrum of density fluctuations \citep[we describe the normalization using $\sigma_8$, the {\it rms} fluctuations of overdensity within spheres of 8$h^{-1}$~Mpc radius; i.e.][]{henry97,viana99, reiprich02}; surveys that probe the cluster population at higher redshift are sensitive to the growth of density fluctuations, allowing one to break the $\Omega_M$-$\sigma_8$ degeneracy that arises from local cluster abundance constraints \cite{eke96,bahcall98}.  Wang \& Steinhardt \cite{wang98} argued that a measurement of the changes of cluster abundance with redshift would provide constraints on the dark energy equation of state parameter $w\equiv p/\rho$.  

Describing the problem in terms of cluster abundance only makes sense in the local universe , because, of course, one cannot measure the cluster abundance without knowing the survey volume; the survey volume beyond $z\sim0.1$ is sensitive to cosmological parameters that affect the expansion history of the universe-- namely, the matter density $\Omega_M$, the dark energy density $\Omega_E$ and the dark energy equation of state $w$.  A cluster survey of a particular piece of the sky with appropriate followup actually delivers a list of clusters with mass estimates and redshifts-- that is, the redshift distribution of galaxy clusters above some detection limit.
Haiman, Mohr \& Holder \cite{haiman01} showed how the redshift distribution from large X--ray or Sunyaev-Zel'dovich effect (SZE) cluster surveys allows for precise measurements of the dark energy equation of state that are competitive with constraints possible from studies of supernovae at high redshift.

A series of papers details recent work  to explore the theoretical and observational obstacles to precise cosmological measurements with cluster surveys.  Holder, Haiman \& Mohr \cite{holder01b} introduced the Fisher matrix formalism and showed that high yield SZE cluster surveys can provide precise constraints on the geometry of the universe through simultaneous measurements of $\Omega_E$ and $\Omega_M$.  Weller, Battye \& Kniessl \cite{weller01} demonstrated that future SZE surveys might constrain the variation of the dark energy equation of state $w(z)$.  Hu \& Kravtsov \cite{hu02} examined the effects of cosmic variance on cluster surveys as well as including the effects of imprecise knowledge of a larger number of cosmological parameters. Levine, Schulz \& White  \cite{levine02} examined an X--ray cluster survey, showing that a sufficiently large survey allows one to measure cosmological parameters and constrain the all--important cluster mass--observable relation simultaneously.  This raises the exciting possibility that cluster surveys are self--calibrating-- that the redshift distribution contains enough energy to solve for cluster structure and cosmology simultaneously!  An important caveat to this work is that the authors assumed that the evolution of cluster structure was perfectly known.  In a more recent work, Majumdar \& Mohr \cite{majumdar02} show that if one allows for our imprecise knowledge cluster structural evolution with redshift, the constraint on the dark energy equation of state $w$ evaporates.  In addition, we show that if one incorporates followup measurements-- perhaps from X--ray, SZE or weak lensing-- into the survey one can recover the precise constraints on $w$.  These calculations underscore the importance of incorporating information from multiple observables into future cluster surveys.  Ongoing calculations by several groups will undoubtedly provide additional insights into how to optimize cluster surveys, but today it appears that high yield cluster surveys are as viable a means of studying dark energy as first suggested.

\section{Prospecting for Clusters}

Galaxy clusters are dark matter dominated objects with baryon reservoirs in the form of an intracluster medum (ICM) and a galaxy population.  Clusters can be found through the light the galaxies emit, the gravitational lensing distortions the cluster mass introduces into the morphologies of background galaxies, the X--rays emitted by the energetic ICM, the distortion that the hot ICM introduces into the cosmic microwave background spectrum (SZE), and the effects that the ICM has on jet structures associated with active galaxies in the cluster.  These methods are largely complementary, each having different strengths.  It appears that X--ray and SZE signatures of clusters are higher contrast observables than are weak lensing or galaxy light.  That is, massive galaxy clusters are more prominent relative to the jumble of lower mass halos and large scale filaments when viewed with the SZE and X-ray;  projection effects are a far more serious concern when using galaxy light or weak lensing signatures.  Studies of the highest redshift galaxy clusters will likely be done with the SZE, because of the redshift independence of the spectral distortion in the CMB.  Further work has to be done to develop optimal means of applying each of these observables and combining them to create the highest fidelity picture of the cluster population and its evolution.

\begin{figure}
\vskip-10pt
\hbox{\hskip-20pt\includegraphics[scale=0.40]{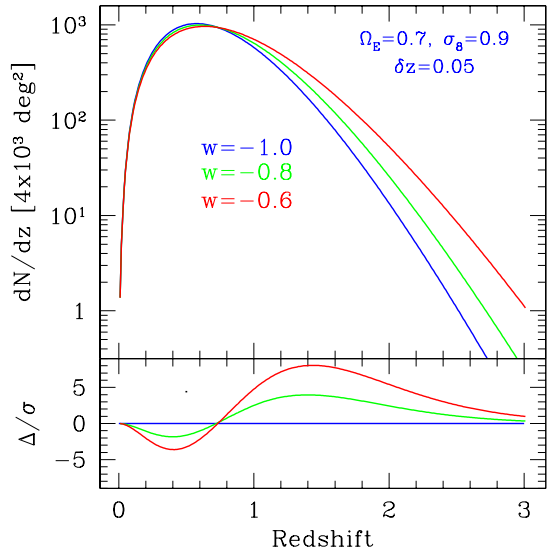}\hfil}
\vskip-40pt
\caption{The cluster redshift distribution for a 4000~deg$^2$ SZE survey proposed for the South Pole is sensitive to the dark energy equation of state $w$.  Three models with different $w$ are shown (above) and statistical differences quantified (below).\label{fig:wsense}}
\vskip-25pt
\end{figure}

\begin{figure}
\vskip-10pt
\hbox{\hskip-20pt\includegraphics[scale=0.40]{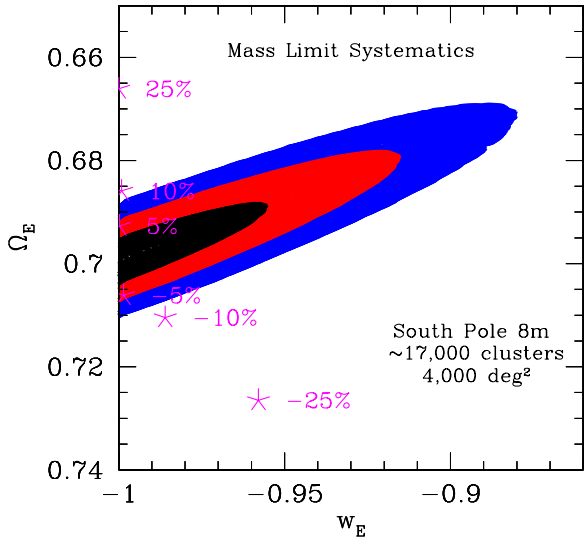}\hfil}
\vskip-40pt
\caption{The South Pole SZE survey should deliver precise constraints on the dark energy density $\Omega_E$ and equation of state $w$.  Statistical confidence regions ($1\sigma$, $2\sigma$ \& $3\sigma$) are shown, and stars mark the parameter biases associated with systematic errors in the cluster mass estimators.  Controlling mass systematics to 10\% or better should allow a $\sim$5\% measurement of $w$.  \label{fig:constrain}}
\vskip-25pt
\end{figure}

Figure~\ref{fig:wsense} contains the redshift distributions for one proposed SZE survey from the South Pole (PI: Carlstrom).  Redshift distributions for $w=-1$, $w=-0.8$ and $w=-0.6$ are shown to demonstrate the effects of this parameter on the survey.  In the lower panel the statistical ``distance'' between these models is shown, illustrating sensitivity near the peak in $dN/dz$ at $z=1/2$ and again at $z=3/2$.  Figure~\ref{fig:constrain} shows estimated 1$\sigma$, 2$\sigma$ and 3$\sigma$ confidence  regions for the survey within the $\Omega_E$--$w$ space.  Only geometrically flat models are considered, and constraints are marginalized over $\sigma_8$.  This figure shows that this large solid angle survey can in principle deliver $\sim$5\% constraints on the nature of dark energy!

\section{Mass--Observable Scaling Relation}

A solid theoretical foundation has emerged for understanding the formation and evolution of massive, dark matter dominated halos \citep[i.e. galaxy clusters;][]{jenkins01,evrard02}.   Using cluster surveys to study cosmology requires that we connect cluster observables to this theoretical foundation.  This connection is made by using cluster mass--observable scaling relations to connect cluster observables like X-ray luminosity, SZE luminosity, weak lensing shear, and galaxy light to cluster halo mass estimates.  Weak lensing offers advantages here, because weak lensing mass estimates don't require that the cluster be in equilibrium \citep[i.e.][]{bartelmann01b}; however, projection of foreground and background mass along the line of sight to a particular cluster can bias lensing mass estimates  \citep{metzler01}.  It should be possible to estimate the scale of these biases using numerical simulations.  

Low redshift clusters do exhibit striking regularity \citep[i.e.][]{david93,mohr97a}, suggesting that observables like the ICM X-ray luminosity and temperature are good mass estimators \citep{finoguenov01,reiprich02}.  Cluster mass to light ratios have been studied for decades, and it may be that this body of work together with modern datasets will allow more conclusive statements about how well galaxy light traces cluster halo mass.  Hydrodynamical simulations lead us to expect that the SZE luminosity (related to the total pressure within the virial region) should be the best ICM observable for predicting mass, but we await next generation SZE instruments and new observations to demonstrate this.  Questions about the persistence of these mass--observable scaling relations to higher redshift where clusters are presumably experiencing more rapid merging remain, but results to date suggest that the degree of cluster regularity locally and at intermediate redshift is comparable \citep{mushotzky97,mohr00a,vikhlinin02}.

Annotated stars in Figure~\ref{fig:constrain} show the effects of systematic mass errors at the $\pm$5\%, 10\% and 25\% level on the cosmological constraints.  In each case the location of the star shows the best fit cosmology, when the input cosmology is $\Omega_E=0.7$ and $w=-1$.  This simple calculation shows the importance of controlling mass--observable relation systematics at the $\sim$10\% level or better in order to fully realize the statistical power of the survey.

\section{Thermal History of the ICM}

The thermal history and enrichment history of the ICM is another critical issue that one can potentially address using galaxy cluster surveys.  The ICM structure of low redshift galaxy clusters as revealed by scaling relations indicates a lack of low entropy gas when compared to simple heirarchical structure formation models.  In addition, the ICM is highly enriched with metals.  These facts suggest that the thermal history of the ICM may be more complex than simple infall models, involving sources of entropy (or preheating) or significant cooling to remove the lowest entropy gas \citep{cavaliere97,bryan00b}.  From a cosmological perspective, this unknown thermal history provides a complication and a possible source of bias in using cluster surveys to measure the nature of dark energy.

Bialek, Evrard \& Mohr \cite{bialek01} have examined the effects of preheating (in our case an initial entropy in the baryon gas that is introduced well before cluster formation commences) on low redshift cluster scaling relations; here we use those same simulations to address the effects that preheating could have on cluster survey cosmology.  We find that an initial entropy of $\sim$100~kev~cm$^2$ is sufficient to match the slopes of the local X--ray luminosity temperature relation \citep[i.e.][]{david93}, the ICM mass--temperature relation \citep{mohr99}, and the X--ray size--temperature relation \citep{mohr97a}.  However, significant uncertainties remain, and there is an ongoing debate about the importance of radiative cooling in changing cluster structure.  Thus, when using surveys to do cosmology, we must consider the initial entropy as an additional variable that could alter the observed redshift distribution.

\begin{figure}
\vskip-10pt
\hbox{\hskip-20pt\includegraphics[scale=0.40]{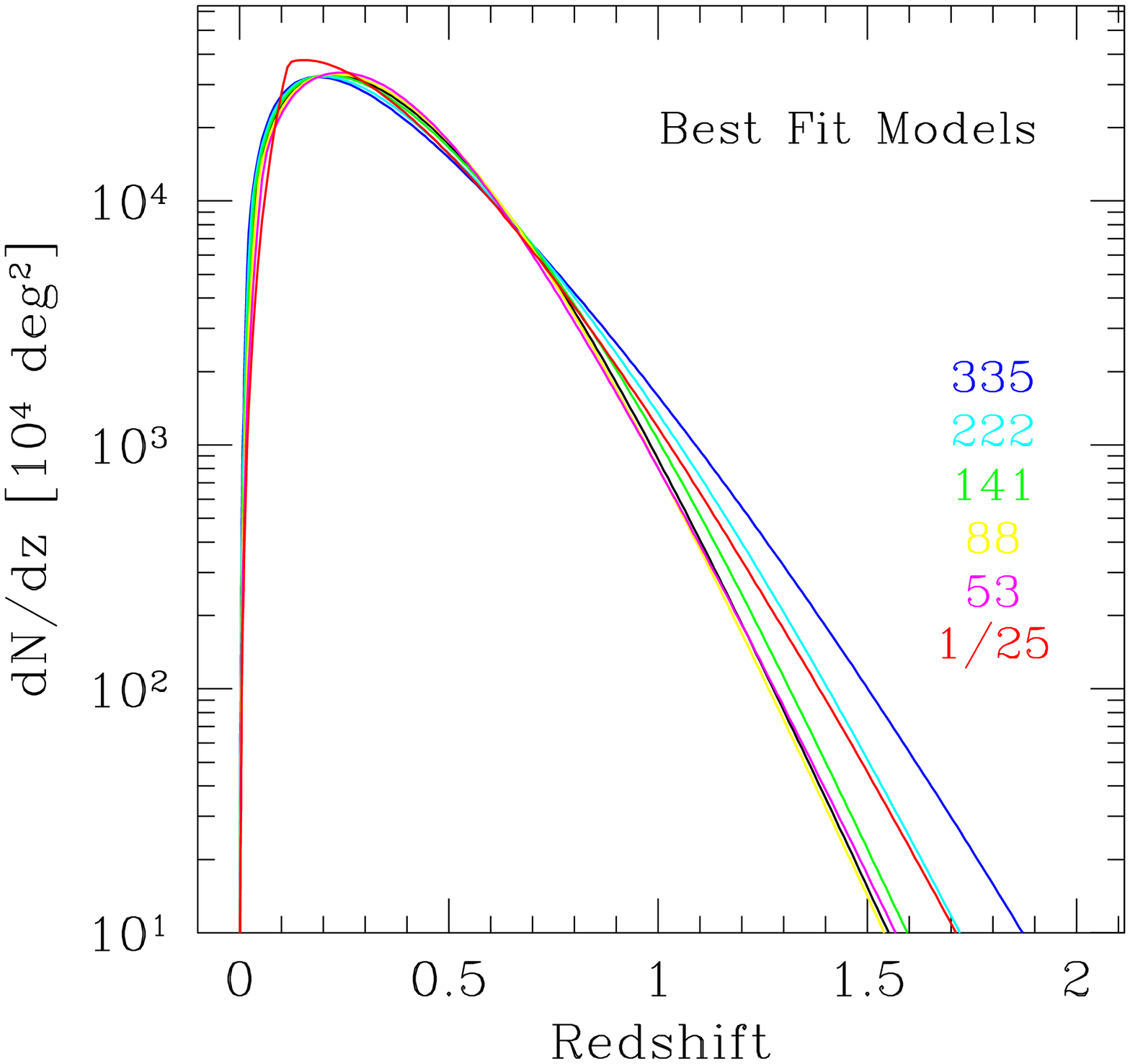}\hfil}
\vskip-25pt
\hbox{\hskip-20pt\includegraphics[scale=0.40]{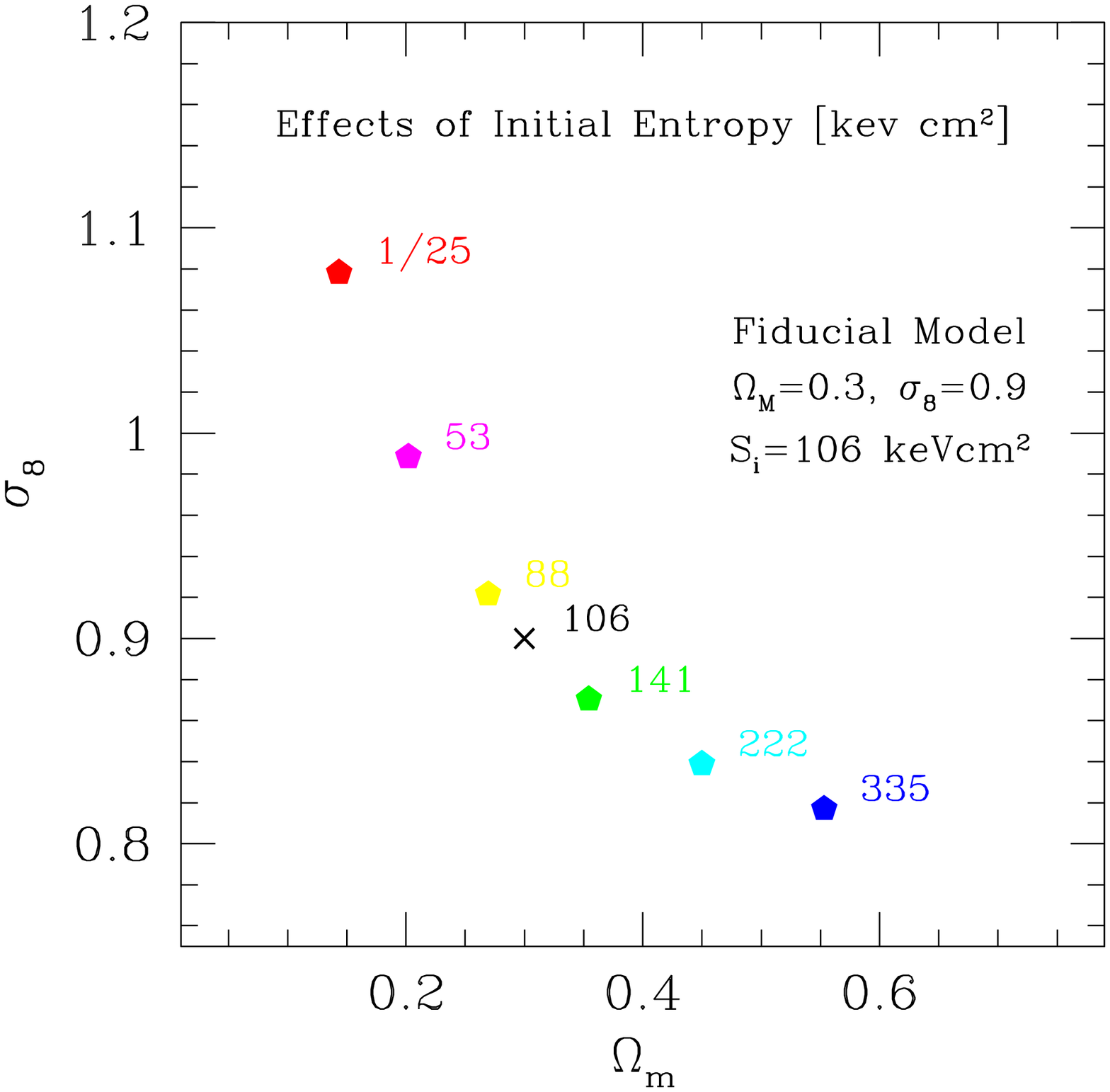}\hfil}
\vskip-40pt
\caption{Cosmological parameter biases associated with imperfect knowledge of the thermal history of the ICM (below) in an X--ray cluster survey.  Annotated points correspond to the best fit models that come when interpreting the survey using the wrong initial entropy (see text).  These best fit models are not ``good'' fits (illustrated above), so X--ray surveys allow one to solve for cosmology and initial entropy simultaneously.  In comparison to X--ray surveys, SZE surveys are essentially unaffected by preheating, suggesting that SZE surveys are poor tools for studies of the thermal history of the ICM.
\label{fig:dndzpre}}
\end{figure}

We examine this issue by assuming a fiducial cosmology with an initial entropy of 106~kev~cm$^2$, $\Omega_M=0.3$ and $\sigma_8=0.9$.  We then interpret an X--ray survey similar to that possible with the DUET Midex mission using different (i.e. wrong) initial entropies.  Figure~\ref{fig:dndzpre} (bottom) shows the best fit parameters that result for a range of initial entropies varying from 1/25 to 335~kev~cm$^2$.  The correct cosmology is recovered when we use an initial entropy of 106~kev~cm$^2$, but $\sigma_8$ is biased by $\sim$10\% and $\Omega_M$ by $\sim$80\% when an incorrect entropy of 335~kev~cm$^2$ is used.  This shows that X-ray surveys are very sensitive to the preheating level, and provide an excellent way of studying the thermal history of the ICM.  This is undoubtedly because much of the X--ray luminosity is emitted from the central cluster region that contains the lowest entropy ICM, whose distribution is most affected by entropy changes.  The top panel shows the best fit $dN/dz$ corresponding to each of the models.  These best fit models are not statistically good fits, and so it is possible to use the survey to determine the correct entropy {\it and} the correct cosmological parameters.

We have used this same approach to examine the sensitivity of SZE surveys to preheating.  Even for high sensitivity SZE surveys that probe the lowest mass clusters, the sensitivity to preheating appears to be extremely low.  This is because preheating introduces offsetting effects that leave the SZE luminosity (related to the total thermal energy within the virial region) almost unchanged:  (i) the ICM temperature is increased somewhat and (ii) the ICM mass fraction is reduced somewhat.  A figure with the same scale as Figure~\ref{fig:dndzpre} but made for an SZE survey would have all best fit models falling indistinguishably close to the correct cosmological model.  SZE surveys are unbiased by our lack of knowledge about the thermal history of the ICM.  The flip side of this is that-- despite some recent claims to the contrary in the literature-- SZE surveys are extremely poor tools for investigating the thermal history of the ICM.

\bibliographystyle{elsart-num}
\def\apj{ApJ}
\def\apjl{ApJ Letters}
\def\aap{Astr \& Ap}
\def\physrep{Physics Reports}
\def\mnras{MNRAS}

\end{document}